\documentclass[onecolumn]{revtex4-1}

\usepackage{amsmath}
\usepackage{amsfonts}
\usepackage{graphicx}
\usepackage{color}

\usepackage[left,modulo]{lineno}
\setcitestyle{authoryear,round}

\begin{document}

\title{Statistical optimization for  passive scalar transport:  maximum entropy production vs maximum Kolmogorov-Sinay entropy}%

\author{M. Mihelich, D. Faranda, B. Dubrulle and D. Paillard}%
\email[REVTeX Support: ]{revtex@aps.org}
\affiliation{Laboratoire SPHYNX, CEA/IRAMIS/SPEC, CNRS URA 2464, F-91191 Gif-sur-Yvette, France;
E-Mail: berengere.dubrulle@cea.fr}
\date{september 16, 2014}

\begin{abstract}
%Ici a garder pour conclusion and abstrait
 We derive rigorous results on the link between the principle of maximum entropy production and the principle of maximum Kolmogorov- Sinai entropy for a Markov model of the passive scalar diffusion called the Zero Range Process. 
We show analytically that  both the entropy production and the Kolmogorov-Sinai entropy, seen as functions of a parameter  $f$ connected to the jump probability, admit a unique maximum denoted $f_{max_{EP}}$ and $f_{max_{KS}}$.  The behavior of these two maxima is explored as a function of the system disequilibrium and the system resolution $N$. The main result of this article is that $f_{max_{EP}}$ and $f_{max_{KS}}$ have the same Taylor expansion at first order in the deviation from equilibrium. We find that $f_{max_{EP}}$  hardly depends on $N$ whereas $f_{max_{KS}}$ depends strongly on $N$.  In particular, for a fixed difference of potential between the reservoirs, $f_{max_{EP}}(N)$ tends towards a non-zero value, while $f_{max_{KS}}(N)$ tends to $0$ when $N$ goes to infinity. For values of $N$ typical of those adopted by Paltridge and climatologists working on MEP ($N \approx 10 \sim 100 $), we show that $f_{max_{EP}}$ and $f_{max_{KS}}$ coincide even far from equilibrium. Finally, we show that one can find an optimal resolution $N_*$ such that $f_{max_{EP}}$ and $f_{max_{KS}}$ coincide, at least up to a second order parameter proportional to the non-equilibrium fluxes imposed to the boundaries. We find that the optimal resolution $N^*$ depends on the non equilibrium fluxes, so that deeper convection should be represented on finer grids. This result points to the inadequacy of using a single grid for representing convection in climate and weather models.  Moreover, the application of this principle to passive scalar transport parametrization is therefore expected to provide both the value of the optimal  flux, and of the optimal number of degrees of freedom (resolution) to describe the system.

\end{abstract}

\maketitle
%\tableofcontents

\section{Introduction}
A major difficulty in the modeling of nonlinear geophysical or astrophysical processes is the taking into account of all the relevant degrees of freedom. For example, fluid motions obeying Navier-Stokes equations usually require of the order of $N=Re^{9/4}$ modes to faithfully describe all
scales between the injection scale and the dissipative scale \citep{frisch1995turbulence}. In atmosphere, or ocean, where the Reynolds number exceeds $10^{9}$, this amount to $N=10^{20}$, a number too large to be handled by any existing computers \citep{wallace2006atmospheric}. The problem is even more vivid in complex systems such as planetary climate, where the coupling of lito-bio-cryo-sphere with ocean and atmosphere increases the number of degrees of freedom beyond any practical figure. This justifies the long historical tradition of parametrization and statistical model reduction, to map the exact equations describing the system onto a set of simpler equations involving few degrees of freedom. The price to pay   is the introduction of \textit{free parameters}, describing the action of discarded degrees of freedom, that needs to be prescribed.\\
 When the number of free parameters is small, their prescription can be  successfully done empirically  through calibrating experiments or by a posteriori tuning \citep{rotstayn2000tuning}. When the number of parameters is large, such as in climate models where it reaches several hundreds \citep{murphy2004quantification}, such empirical procedure is inapplicable, because it is impossible to explore the whole parameter space. In that respect, it is of great interest to explore  an alternative road to parametrization via application of a statistical optimization
principle, such as minimizing or maximizing of a suitable cost functional. As discussed by \citep{turkington2013optimization} and \citep{pascale2012parametric}, this strategy usually leads to 
closed reduced equations with adjustable parameters in the closure appearing as weights in
the cost functional and can be computed explicitly. A famous example in climate is given by a principle of maximum entropy production (MEP)
that allowed  \citep{paltridge1975global} to derive the distribution of heat and clouds at the Earth surface with reasonable accuracy, without any parameters and with a model of a dozen of degrees of freedom (boxes). Since then, refinements of Paltrige model have been suggested to increase its generality  and range of prediction \citep{herbert2011present}. MEP states that a stationary nonequilibrium system chooses its final state in order to maximize the entropy production as is explain in \citep{martyushev2006maximum}. Rigorous justifications of its application have been searched using e.g. information theory \citep{dewar2014theoretical} without convincing success. More recently, we have used the analogy of the climate box model of Paltridge with the asymmetric exclusion Markov process (ASEP) to establish numerically a link between the MEP and
the principle of maximum Kolmogorov- Sinai entropy (MKS)\citep{mihelich2014maximum}. The MKS principle is a relatively new concept which extends the classical results of equilibrium physics \citep{monthus2011non}. This principle applied to Markov Chains  provides an approximation of the optimal diffusion coefficient in transport phenomena \citep{gomez2008entropy} or simulates random walk on irregular lattices \citep{burda2009localization}. It is therefore a good candidate for a physically relevant cost functional in passive scalar modeling.\\

The goal of the present paper is to derive rigorous results on the link between MEP and MKS using a Markov model of the passive scalar diffusion called the Zero Range Process \citep{andjel1982invariant}. We find that there exists an optimal resolution $N_*$ such that both maxima coincide to second order in the distance from equilibrium. The application of this principle to passive scalar transport parametrization is therefore expected to provide both the value of the optimal  flux, and of the optimal number of degrees of freedom (resolution) to describe the system. This suggests that the MEP and MKS principle may be unified when the  Kolmogorov- Sinai entropy  is defined on opportunely coarse grained partitions. 

\section{From passive scalar equation to ZRP model}
The equation describing the transport of a passive scalar like temperature in a given velocity field $u(x,t)$ reads:
\begin{equation}
\label{eq:diff0}
\partial_tT+u\partial_xT=\kappa  \partial_x^2 T,
\end{equation}

with appropriate boundary conditions, or equivalently,  in non-dimensionnal form: \begin{equation} \partial_t T+u\partial_x T=\frac{1}{RePr}\partial_x^2 T, \end{equation} where  $\kappa$, $Re$ and $Pr$ are respectively the molecular diffusivity, the Reynolds and the Prandtl number. To solve this equation, one must know both the velocity field and the boundary conditions, and use as many number of modes as necessary to describe all range of scales up to the scales at which molecular diffusivity takes place i.e. roughly
$(Re Pr)^{3/2}$ modes, where $Re$ is the Reynolds number of the convective flow, and $Pr$ is its Prandtl number. In geophysical flows, this number is too large to be handled even numerically \citep{troen1986simple}. Moreover, in typical climate studies, the velocity flow is basically unknown as it must obey a complicated equation involving the influence of all the relevant climate components. In order  to solve the equation, one must necessarily  prescribe the heat flux $f=-uT+\kappa \nabla T$. The idea of Paltridge was then to discretize the passive scalar equation in boxes and prescribe the heat flux $f_{i(i+1)}$ between boxes $i$ and $i+1$  by maximizing the associated thermodynamic entropy production $\dot S=\sum_i f_{i(i+1)}(\frac{1}{T_{i+1}}-\frac{1}{T_{i}})$. 
Here, we  slightly modify the Paltridge discretization approximation to make it amenable to rigorous mathematical results on Markov Chains.
For simplicity, we stick to a one dimensional case (corresponding to boxes varying only in latitude) and impose the boundary conditions through two reservoirs located at each end of the chain (mimicking the solar heat flux at pole and equator).  We consider a set of \(N\) boxes that can contain an arbitrary number  \(n \in \mathbb{N}\) of particles. We then allow transfer of particles in between two adjacent boxes via 
decorrelated jumps (to the right or to the left) following a 1D Markov dynamics governed by a coupling with the two reservoirs imposing a difference of chemical potential at the ends. The resulting process is called the Zero Range Process \citep{andjel1982invariant}. 
The different jumps are described as follow. At each time step a particle can jump right with probability \(pw_n\) or jump left with probability \(qw_n\) where \(w_n\) is a parameter depending of the number of particles  inside the box. Physically it represents the interactions between particles. At the edges of the lattice the probability rules are different:  At the left edge a particule can enter with probability \(\alpha\) and exit with probability \(\gamma w_n\) whereas at the right edge a particle can exit with probability \(\beta w_n\) and enter with probability \( \delta\). Choices of different $ w_n$ give radically different behaviors. For example $ w_n = 1+ b/n$ where $b\geq 0$ described condensation phenomena \citep{grosskinsky2003condensation} whereas $ w_1=w$ et $ w_n=1$ if $n\geq 2$ has been used to modeled road traffic. We will consider in this article the particular case where \(w=1\) by convenience of calculation. Moreover without loss of generality we will take \(p \geq q\) which corresponds to a particle flow from the left to the right and note  \(f=p-q\)  . After a sufficiently long time the system reaches a non-equilibrium steady state. The interest of this toy model is that it is simple enough so that exact computations are analytically tractable.\\

Taking the continuous limit of this process, it may be checked that the fugacity $z$, which  is a quantity related to the average particle
density (see \ref{eq:3} below),  of stationary solutions of a system consisting of boxes of size $\frac {1}{N}$ follows the continuous equation  \citep{levine2005zero} :
\begin{equation}
\label{eq:diff1}
f\frac{\partial z}{\partial x}- \frac{1}{2N}\frac{\partial^2 z}{\partial x^2}=0,
\end{equation}

corresponding to a stationary solution of a non-dimensional passive scalar equation with non-dimensionnal velocity $f$ and a non-dimensionnal diffusivity $\frac{1}{RePr}=\frac{1}{2N}$. Therefore, the fugacity of the Zero Range Process is a passive scalar obeying a convective-diffusion equation, with advection velocity controlled by the probability to jump to the right or to the left, and diffusivity controlled by the number of boxes: the larger the number of boxes (the finer the resolution) the smaller the diffusivity. This observation illuminates the well-known observation that the numerical diffusion of a discrete model of diffusion is inversely proportionnal to the resolution. The parameter $f$  controls the regime: $f=0$ corresponds to
 a purely conductive regime whereas the larger $f$ the more convective the regime. In the sequel, we calculate the entropy production and the Kolmogorov-Sinai entropy function of $f$. These two quantities reach a maximum noted respectively \(f_{max_{EP}}\) and $ f_{max_{KS}} $. The MEP  principle (resp. the MKS principle) states that the system will choose $f=f_{max_{EP}}$ (resp $f=f_{max_{KS}}$). \\
We  will show first of  all  in this article that numerically \(f_{max_{EP}} \approx f_{max_{KS}} \)   even far from equilibrium for a number of boxes $N$  roughly corresponding to the resolution taken by  \cite{paltridge1975global} in his climate model. This result is similar to what we found for the ASEP model \citep{mihelich2014maximum} and thus gives another example of a system in which the two principles are equivalent. Moreover we will see analytically that  \(f_{max_{EP}}\) and $ f_{max_{KS}} $ have the same behavior in first order in the difference of the chemical potentials between the two reservoirs for $N$ large enough.  These results provide a better understanding of the relationship between the MEP and the MKS principles. \ \

\section{Notations and useful preliminary results}

This Markovian Process is a stochastic process with a infinite number of states in bijection with  $ \mathbb{N}^N$. In fact, each state can be written  $ n=(n_1,n_2,....,n_N)$ where $n_i$ is the number of particule lying in site $i$. We call \(P_n\) the stationary probability  to be in state $n$. In order to calculate this probability it is easier to use  a quantum formalism than the Markovian formalism as explained in the following articles \citep{domb2000phase,levine2005zero}.\\

%\textbf{ EXPLIQUER UN PEU LE FORMALISME QUANTIQUE PEUT ETRE}

The probability to find $m$ particles in the site $k$ is equal to: $ p_k(n_k=m)=\frac{z_k^m}{Z_k}$ where $Z_k$ is the analogue of the grand canonical repartition function  and $z_k$ is the fugacity between $0$ and $1$. Moreover $Z_k = \sum_{i=0}^{\infty} z_k^i=\frac{1}{1-z_k}$. So, finally

\begin{equation}
\label{eq:proba0}
p_k(n_k=m)=(1-z_k) z_k^m,
\end{equation}

We can show that the probability $P$ over the states is the tensorial product of the probability $p_k$ over the boxes:

\[P=p_1 \otimes p_2 \otimes ....\otimes p_N,\] 

Thus events $ (n_k=m)$ and $(n_k'=m')$ for $ k\neq k'$ are independent and so:

\begin{equation}
\label{eq:proba1}
P(m_1,m_2,...,m_N)=p_1(n_1=m_1)*...*p_N(n_N=m_N),
\end{equation}

So finally

\begin{equation}
\label{eq:proba2}
 P(m_1,m_2,...,m_N) = \prod_{k=1}^{N} (1-z_k)z_k^{m_k}.
\end{equation}

Moreover, with the Hamiltonian equation found from the quantum formalism we can find the exact values of $z_k$ function of the system parameters:

\begin{equation}
\label{eq:1}
z_k=\frac{ (\frac{p}{q})^{k-1}[(\alpha+\delta)(p-q)-\alpha \beta + \gamma \delta]-\gamma \delta +\alpha \beta (\frac{p}{q})^{N-1}}{\gamma(p-q-\beta)+\beta(p-q+\gamma)(\frac{p}{q})^{N-1}},
\end{equation}

and the flux of particles $c$:

\begin{equation}
\label{eq:2}
c=(p-q) \frac{-\gamma\delta+\alpha\beta(\frac{p}{q})^{N-1}}{\gamma(p-q-\beta)+\beta(p-q+\gamma)(\frac{p}{q})^{N-1}}.
\end{equation}

Finally, the stationary density is related to the fugacity by the relation:

\begin{equation}
\label{eq:3}
\rho_k=z_k  \frac{\partial \log Z_k}{\partial z_k}=\frac{z_k}{1-z_k}.
\end{equation}

\subsection{ Entropy Production }

For a system subject to internal forces \(X_i\) and associated fluxes \(J_i\) the macroscopic entropy production is well known \citep{onsager1931reciprocal} and takes the form:

\[ \sigma=\sum_{i} J_i*X_i.\]

The Physical meaning of this quantity is a measure of irreversibility: the larger \(\sigma\)  the more irreversible the system.

In the case of the zero range process irreversibility is created by the fact that \(p\neq q\). We will parametrize this irreversibility by the parameter $f=p-q$ and we will take $ p+q=1$.  In the remaining of the paper, we take, without loss of generality, $p \leq q$ which corresponds to a flow from left to right. Moreover, the only flux to be considered is here the flux of particules  $c$ and the associated force is due to the gradient of the density of particules $\rho$ : $X= \nabla \log \rho$ \citep{balian}.

Thus, when the stationary state is reached ie when \(c\) is constant:

\begin{equation}
\label{eq:station}
\sigma=\sum_{i=1}^{N-1} c.(\log(\rho_i)-\log(\rho_{i+1}))=c.( \log(\rho_1)-log(\rho_N)).
\end{equation}

Thus, according to Eqs. (\ref{eq:1}), (\ref{eq:2}),  (\ref{eq:3}) and (\ref{eq:station})  when $ N$ tends to $ + \infty $ we obtain:

\begin{equation}
\label{eq:4}
\sigma(f) = \frac{\alpha f}{ f+\gamma}(\log(\frac{\alpha}{f+\gamma-\alpha})-\log(\frac{(\alpha+\delta)f+\gamma\delta}{f(\beta-\alpha-\delta)+\beta\gamma-\gamma\delta})).
\end{equation}

 Because $f \geq 0$ the entropy production is positive if and only if $ \rho_1 \geq \rho_N$ iff $z_1 \geq z_N$. This is physically coherent because fluxes are in the opposite direction of the gradient.  We remark that if $f=0$ then $\sigma(f)=0$. Moreover, when $f$ increases $\rho_1(f)$ decreases and $ \rho_2(f) $ increases till they take the same value. Thus it exists $f$, large enough, for which $\sigma(f)=0$. Between these two values of $f$ the entropy production has at least one maximum.

\subsection{ Kolmogorov-Sinai Entropy}

There are several ways to introduce the Kolmogorov-Sinai entropy which is a mathematical quantity introduced by Kolmogorov and developed by famous mathematician  as  Sinai and Billingsley \citep{billingsley1965ergodic}. Nevertheless, for a Markov process we can give it a simple physical interpretation: the Kolmogorov-Sinai entropy is the time derivative of the Jaynes entropy (entropy over the path).

\begin{equation}
\label{eq:8}
S_{Jaynes}(t)= - \sum_{\Gamma_{[0,t]}} p_{\Gamma_{[0,t]}}.\log(p_{\Gamma_{[0,t]}}),
\end{equation}

For a Markov Chain we have thus:

\begin{equation}
\label{eq:jaynes}
S_{Jaynes}(t)-S_{Jaynes}(t-1)=-\sum_{(i,j)} \mu_{i_{stat}}p_{ij}\log(p_{ij}),
\end{equation}

where $ \mu_{stat}=\mu_{i_{stat}} i=1...N$ is the stationary measure and where the $ p_{ij}$ are the transition probabilities.

Thus the Kolmogorov-Sinai entropy takes the following form:
\begin{equation}
\label{eq:9}
h_{KS}=-\sum_{(i,j)} \mu_{i_{stat}}p_{ij}\log(p_{ij}),
\end{equation}

For the Zero Range Process ,we show in appendix that it can be written as:
\begin{multline}
\label{eq:hks2}
h_{KS}= -(\alpha \log \alpha +\delta \log \delta  +\gamma \log \gamma +\beta \log \beta + (N-1)(p\log(p)+q\log(q)))
+(p\log(p)+q\log(q)) \sum_{i=1}^{N} (1-z_i) \\
+(\gamma\log(\gamma)+p\log(p))(1-z_1) 
+(\beta\log(\beta)+q\log(q))(1-z_N). \\
\end{multline}

 \section{Results}

We will start first by pointing to some interesting properties of $f_{max_{EP}}$ and $f_{max_{KS}}$, then by presenting numerical experiments on the ZRP model and finally concluding with some analytical computations. 

Let us  first note that for $N$,$\alpha$,$\beta$,$\gamma$,$\delta$ fixed the entropy production as well as the Kolmogorov-Sinai entropy seen as functions of  $f$ admit both a unique maximum. 
When $N$ tends to infinity and $f=0$, using  Eq.(\ref{eq:1}) (i.e. the symmetric case), we find  that $z_1=\frac{\alpha}{\gamma}$  and $z_N=\frac{\delta}{\beta}$ . Thus, the system is coupled with two reservoirs with respective chemical potential  $\frac{\alpha}{\gamma}$(left) and   $\frac{\delta}{\beta}$ (right). For $\frac{\alpha}{\gamma} \neq \frac{\delta}{\beta}$ the system is   out of equilibrium. We assume, without loss of generality, $z_1 \geq z_N$ which corresponds to a flow from left to right.  As a measure of deviation from equilibrium we take  $ s= z_1-z_N$: the larger $s$, the more density fluxes we expect into the system. 

First we remark that $f_{max_{EP}}$   hardly depends on $N$ whereas $f_{max_{KS}}$ depends strongly on $N$. This is easily understood because $\sigma$ depends only on $z_1$ and $z_N$ whereas $h_{KS}$ depends on all the  $z_i$. Moreover, the profile of the $z_i$ depends strongly on $N$.  In particular, for a fixed difference of potential between the reservoirs , $f_{max_{EP}}(N)$ tends towards a non-zero value, while $f_{max_{KS}}(N)$ tends to $0$ when $N$ goes to infinity.

Moreover, $f_{max_{EP}}$ and $f_{max_{KS}}$ coincide even far from equilibrium for $N$ corresponding to the choice of  \cite{paltridge1975global} $N \approx 10 \sim 100 $.  For $N$ fixed, as large as one wants, and for all $\epsilon$, as small as one wants, it exists $\nu$  such that for all $s \in [0;\nu]$ $ |f_{max_{EP}}-f_{max_{KS}}| \leq \epsilon$.

\begin{figure}[hb!]
   \begin{minipage}{0.79\columnwidth}
 \includegraphics[width=0.79\columnwidth]{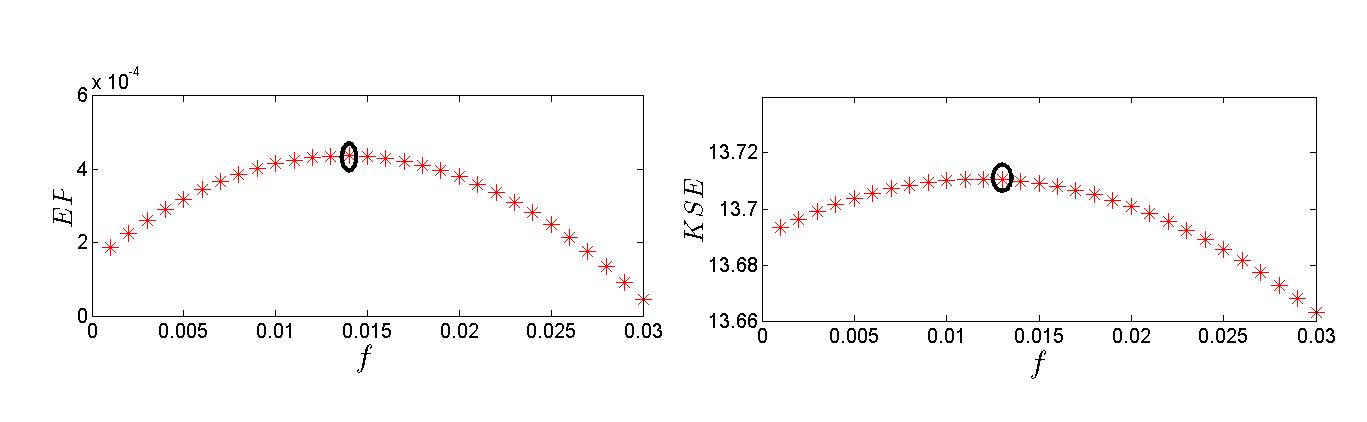}
 \includegraphics[width=0.79\columnwidth]{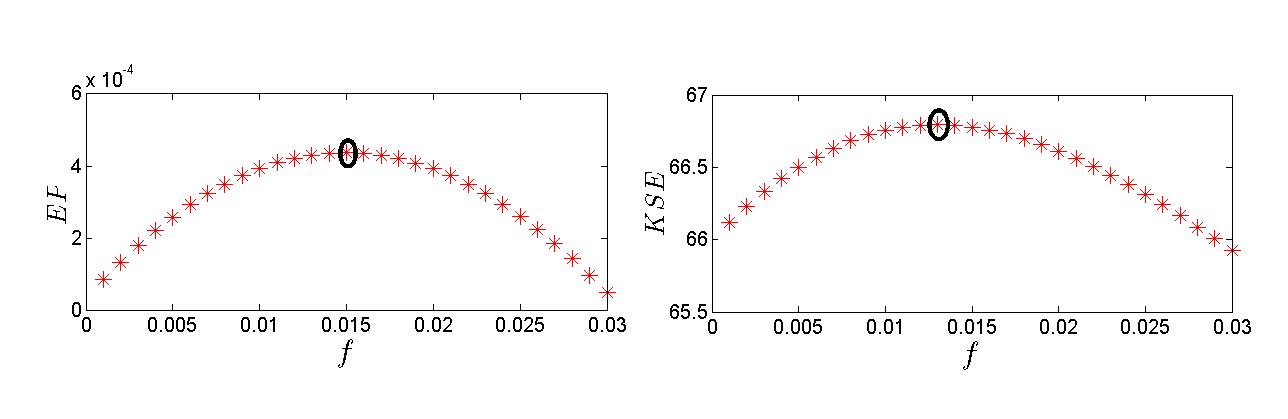}
 \includegraphics[width=0.79\columnwidth]{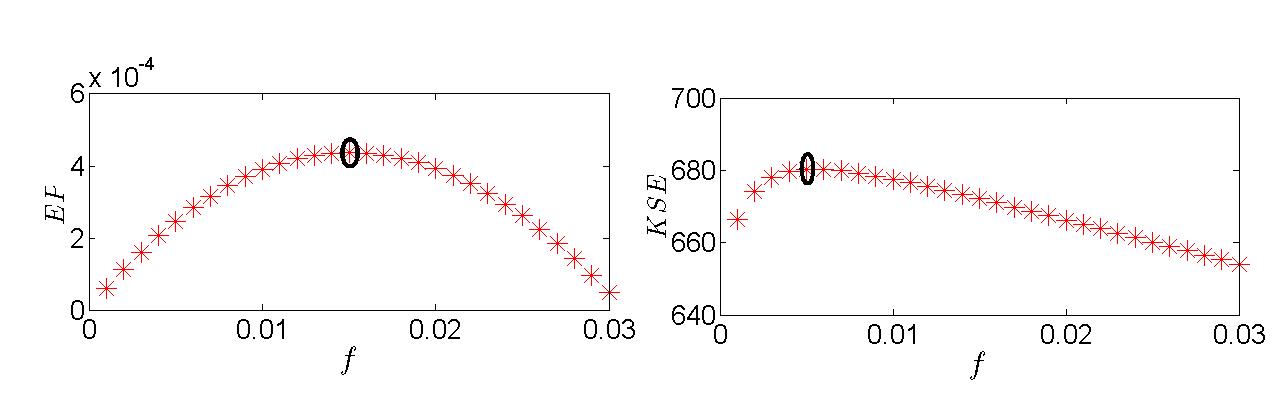}

\end{minipage}
  \caption{ Entropy Production calculate using \ref{eq:4} (left) and KS Entropy calculate using \ref{eq:1} and \ref{eq:hks2} (right) function of $f$  for $s=0.13$ and respectively $N=20$ $N=100$ et $N=1000$}
 \label{difference1}
 \end{figure}

\begin{figure}[hb!]
   \begin{minipage}{0.99\columnwidth}
 \includegraphics[width=0.65\columnwidth]{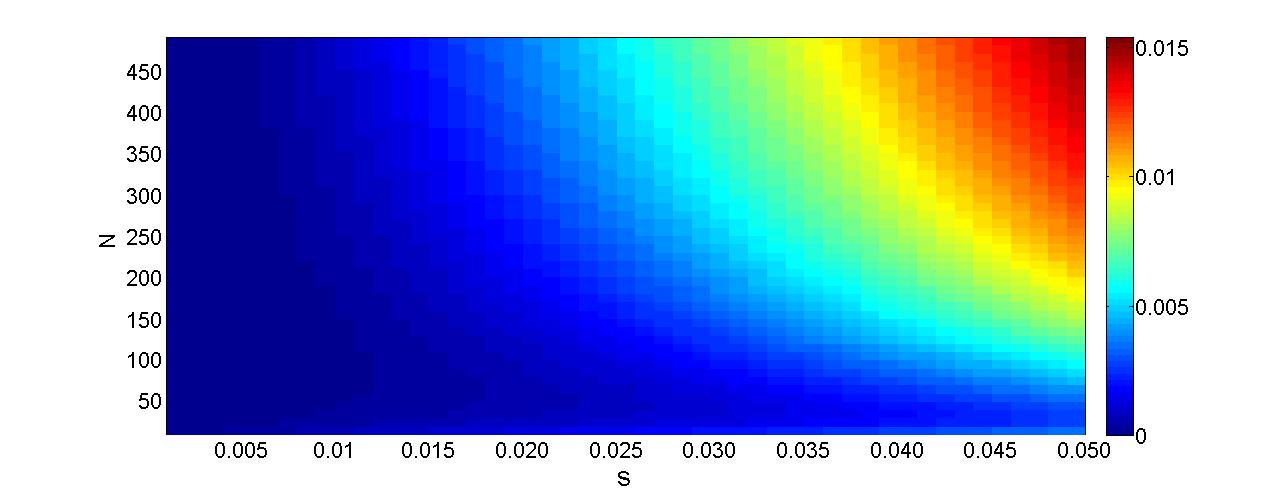}
\end{minipage}
  \caption{   2D plot representing $\Delta f_{max}=f_{max_{EP}}-f_{max_{KS}}$ in the $ (N,s)$ space. }
  \label{2D}
 \end{figure}

These observations are confirmed by the results presented in Figures \ref{difference1} and \ref{2D} where  EP and KS are calculated using Eq.  (\ref{eq:1}) and (\ref{eq:hks2}) for $s=0.13$ and three different partitions: $N=20$ $N=100$ et $N=1000$. The figure  shows that $f_{max_{EP}}$ and $f_{max_{KS}}$ coincide with good approximation for $N=20$ and $N=100$. But then when $N$ increases $f_{max_{KS}}(N)$ tends to $0$ whereas $f_{max_{EP}}(N)$ tends to a non-zero value.

\begin{figure}[hb!]
   \begin{minipage}{0.79\columnwidth}
 \includegraphics[width=0.79\columnwidth]{alphagamma3.jpg}
 \includegraphics[width=0.79\columnwidth]{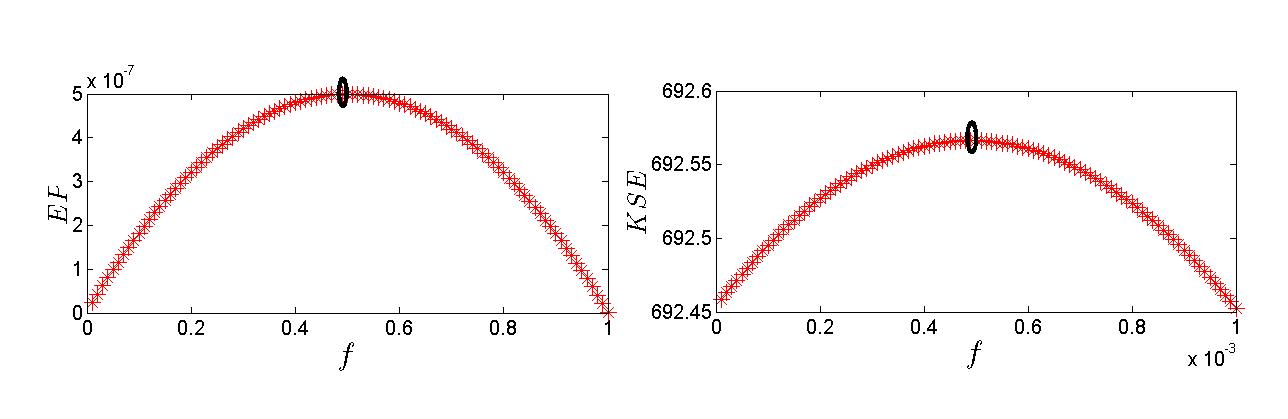}
 \includegraphics[width=0.79\columnwidth]{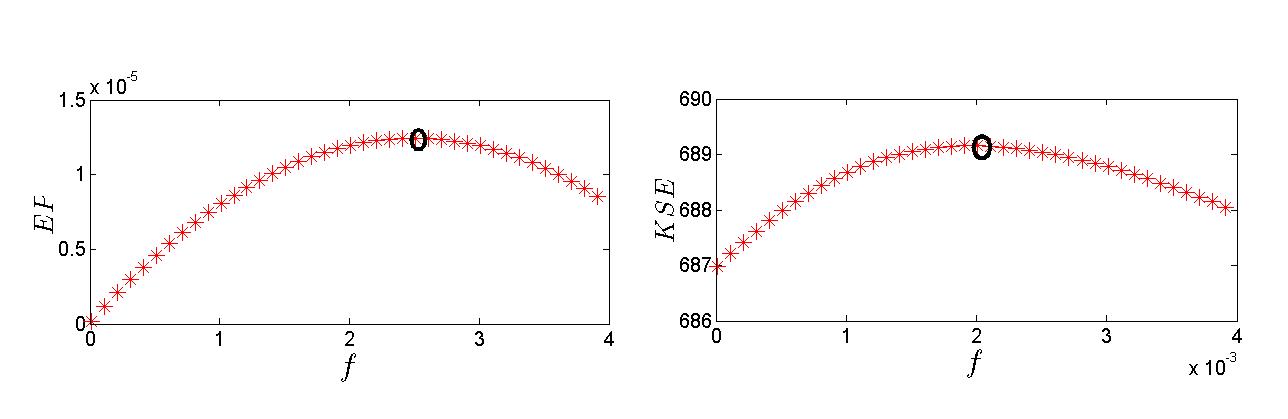}

\end{minipage}
  \caption{ Entropy Production (left) and KS Entropy (right) function of $f$  for $N=1000$ and respectively $s=0.13$; $s=0.2$; $s=0.04$}
  \label{difference2}
 \end{figure}

In Figure \ref{difference2}  we represent  the  Entropy Production (top) and KS Entropy (bottom) function of $f$  for $N=1000$ and  for  three value of $s$:  $s=0.13$; $s=0.2$; $s=0.04$. This supports the claim that for $N$ fixed, we  could tried different values of $s$ such that  $s \in [0;\nu]$ $ |f_{max_{EP}}-f_{max_{KS}}| \leq \epsilon$. Figure \ref{2D} shows that $\Delta f_{max}$ is minimum when the system is close to equilibrium whereas the further the system is from equilibrium (when $s$ increases)  the more $\Delta f_{max}$ increases. Moreover the optimal resolution where $f_{max_{EP}}$ and $f_{max_{KS}}$ coincide  is approximately $10 \sim 100$. Then $\Delta f_{max}$ is maximum at $N =500$ and $s =0.05$. $\Delta f_{max}$  is obvously linear in $s$, for small values of $s$, but the behaviour with $N$  is more complicated.

Such numerical investigations suggest  to understand why $f_{max_{KS}}(N)$ and  $f_{max_{EP}}(N)$ have different behaviors function of $N$, and why for $N$ large enough $f_{max_{KS}}$ and  $f_{max_{EP}}$  have the same behavior of first order in the deviation from equilibrium measured by the parameter $s$.
We will see that we can get a precise answer to such questions by doing calculations and introducing a sort of Hydrodynamics approximation.

\subsection{Taylor expansion}

From Eq. (\ref{eq:hks2}) it is apparent that $f_{max_{KS}}$ depends on $N$  whereas from Eq. (\ref {eq:station}) we get that $f_{max_{EP}}$  hardly depends on $N$.  Indeed there is a difference between $f_{max_{EP}}$ and $f_{max_{KS}}$, i.e. a difference between the two principles for the Zero Range Process.  Nevertheless, we have seen numerically that there is a range of $N$, namely $N \approx 10 \sim 100$  for which the maxima fairly coincide.

Using Eqs.  (\ref{eq:hks2}) (\ref{eq:1}) (\ref{eq:4})  we compute analytically the Taylor expansion of $f_{max_{EP}}$ and $f_{max_{KS}}$ in $s$. We will show the main result: $f_{max_{EP}}$ and $f_{max_{KS}}$  have the same Taylor expansion in first order in $s$ for $N$ large enough. Their Taylor expansions are different up to the second order in $s$ but it exists an $N$, \textit{i.e.} a resolution, such that $f_{max_{EP}}$ and $f_{max_{KS}}$ coincident up to the second order.

Let us start by computing $f_{max_{KS}}$. It does  not depend of the constant terms of  $h_{KS}$ in Eq.(\ref{eq:hks2}) and therefore we need only concern ourselves with :

\begin{equation}
-(p\log(p)+q\log(q)) (\sum_{i=1}^{N} (z_i) -1)+(\gamma\log(\gamma)+p\log(p))(1-z_1) + (\beta\log(\beta)+q\log(q))(1-z_N)=N.H(f,N,\alpha,\gamma,\beta,\delta).
\end{equation}

Using Eq.(\ref{eq:1}), the expression of $H(f,N,\alpha,\gamma,\beta,\delta)$ takes an easy form. To simplify the calculations, we restrict the space of parameter by assuming $ \alpha+\gamma=1$ and $\beta+\delta=1$ and we parametrize the deviation from equilibrium by the parameter $ \bar{s}=\alpha-\delta$. Moreover let's note $a=\frac{1}{N}$. Thus, we have
$ H(f,N,\alpha,\gamma,\beta,\delta)=H(f,a,\alpha,\bar{s})$. In order to know the Taylor expansion to the first order in $\bar{s}$ of $f_{max_{KS}}$ we develop $ H(f,a,\alpha,\bar{s})$ up to the second order in $f$; \textit{i.e.} we have
$ H(f,a,\alpha,\bar{s}) =C+Bf+Af^2 +o(f^2)$ then we find $f_{max_{KS}}=-B/2A$ that we will develop in power of $\bar{s}$. This is consistent if we assume $  f \ll a$.

After some tedious but straightforward calculations, we get at the first order in $\bar{s}$

\begin{equation}
f_{max_{KS}}(\bar{s})=\frac{1}{4} \frac{(1-\alpha)-a(\alpha+2)}{\alpha(1-\alpha)+2a\alpha(\alpha-1)}\bar{s} +o(\bar{s}).
\end{equation}

and so,

\begin{equation}
f_{max_{KS}}(\bar{s})=\frac{1}{4\alpha}\bar{s} +\frac{3a}{4(\alpha-1)}\bar{s}+o(\bar{s}) + o(a\bar{s}).
\end{equation}

 We repeat the same procedure starting from Eq.(\ref{eq:4}) and we obtain:

\begin{equation}
f_{max_{EP}}(\bar{s})=\frac{\bar{s}}{4\alpha} +o(\bar{s})+o(a).
\end{equation}

Thus, since $a=\frac{1}{N} \ll1$ the behaviour of $ f_{max_{KS}}(\bar{s})$ and $f_{max_{EP}}(\bar{s})$  is the same for  $\bar{s}$ small enough.

We remark that we can strictly find the same result by solving the hydrodynamics continuous approximation given by Eq. (\ref{eq:diff1}). This equation is a classical convection-diffusion equation. We remark that, by varying $f$, we change the convective behavior: $f=0$ corresponds to a purely diffusive regime whereas by increasing $f$ we enhance the role of convection. If the system is near equilibrium then $f_{max_{EP}}\approx f_{max_{KS}} \approx 0$ and the system is purely diffusive. When the system is out of equilibrium $f_{max_{EP}}$ and $f_{max_{KS}}$ are different from $0$ and corresponds to an (optimal) trade-off between purely diffusive and convective behavior.

One can verify this numerically: We first calculate the exact values of the Entropy Production function of $f$ using Eq. (\ref{eq:1}) and the Kolmogorov-Sinai Entropy function of $f$ using Eqs. (\ref{eq:1})  (\ref{eq:hks2}). Then we approximate these two curves with a cubic spline approximation in order to find $f_{max_{EP}}$  and $ f_{max_{KS}}$ .

\begin{figure}[hb!]
   \begin{minipage}{0.99\columnwidth}
 \includegraphics[width=0.99\columnwidth]{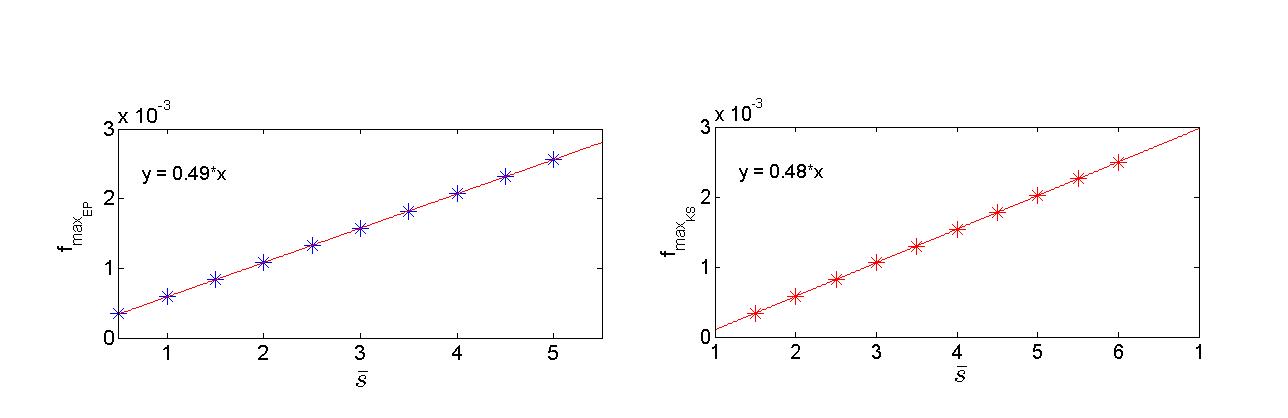}
\end{minipage}
  \caption{  $f_{max_{EP}}$ (left) and $ f_{max_{KS}}$ (right) function of $\bar{s}$ for $\alpha=0.5$ and $N=100$. We remark than $ f_{max_{KS}}$ and $f_{max_{EP}}$ have both a linear behaviour with slope respectively $0.48$ and $0.49$ which is really close to $ \frac{1}{4\alpha}=0.5$ }
  \label{difference3}
 \end{figure}

\begin{figure}[hb!]
   \begin{minipage}{0.99\columnwidth}
 \includegraphics[width=0.99\columnwidth]{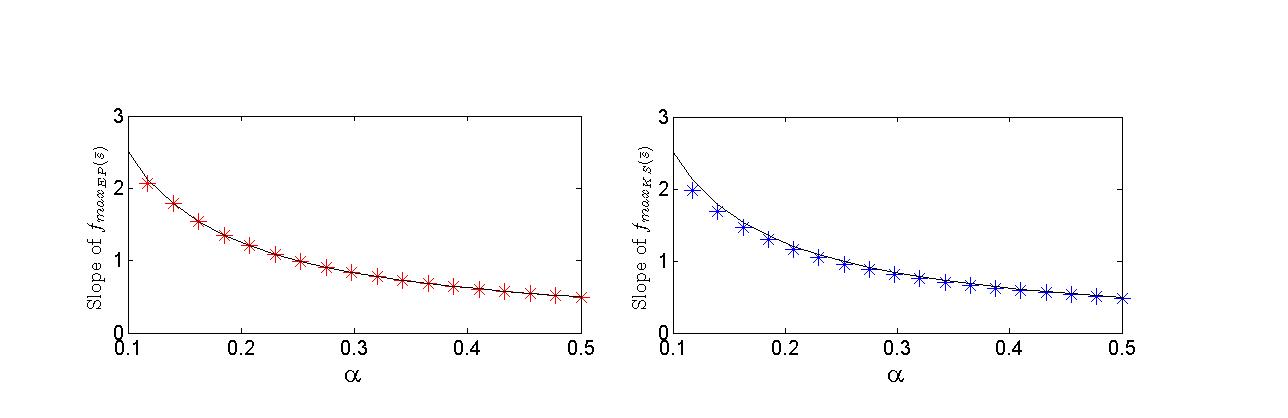}
\end{minipage}
  \caption{ We plot the slope of  $ f_{max_{KS}(\bar{s})}$ (left) and $f_{max_{EP}(\bar{s})}$ (right) function of $\alpha$ and in black the curve $ f(\bar{s})=\frac{1}{4\alpha}\bar{s}$. We remark than the approximation $  f_{max_{KS}}(\bar{s}) \approx  f_{max_{EP}}(\bar{s}) \approx = \frac{1}{4\alpha} \bar{s}$ is good }
  \label{difference4}
 \end{figure}

In order to find the optimal resolution $N_*$ we can go one step further by expanding $f_{max_{EP}}$  and $ f_{max_{KS}}$  up to the second order in $\bar{s}$:

\begin{equation}
f_{max_{EP}}(\bar{s})=\frac{\bar{s}}{4\alpha} +\frac{\bar{s}^2(\alpha+1)}{8\alpha^2(\alpha-1)} +o(\bar{s^2})+o(a).
\end{equation}

\begin{equation}
f_{max_{KS}}(\bar{s})= \frac{1}{4} \frac{(1-\alpha)-a(\alpha+2)}{\alpha(1-\alpha)+2a\alpha(\alpha-1)}\bar{s}+\frac{(1-\alpha)^2+a(\alpha^2-2\alpha+1)}{8\alpha^2(\alpha-1)^2(1-2a)}\bar{s}^2  + o(\bar{s}^2).
\end{equation}

Thus, $f_{max_{EP}}$  and $ f_{max_{KS}}$ coincide in second order in $\bar{s}$ iff $a$ satisfies the quadratic equation:\begin{equation}
(4\alpha-6\alpha^2+6\alpha^3-4\bar{s}+3\alpha^2\bar{s})a^2 -\frac{1}{2}(8\alpha-8\bar{s}+3\alpha^2\bar{s}-6\alpha^2+6\alpha^3)a - (1-\alpha)=0.
\end{equation}

This equation has a unique positive solution because the leading coefficient is positive for $s$ small enough $  (4\alpha-6\alpha^2+6\alpha^3-4\bar{s}+3\alpha^2\bar{s}) \geq 0$ and the constant term is negative  $ - (1-\alpha) \leq 0$. We  remark that the optimal resolution $N_*=\frac{1}{a*}$ depends on the parameters of the system namely on the  degree of non-equilibrium. This fact can be the explanation for two well known issues in climate/weather modeling. First, it explains that, when downgrading or upgrading  the resolution of  convection models, the relevant parameters must be changed as they depend on the  grid size. Second, it suggests that if the   resolution is well tuned to represent a particular range of convective phenomena, it might fail in capturing the dynamics out of this range: since finer grids are needed to better represent  deep convection phenomena, the deviations between model and observations observed in the distribution of extreme convective precipitation  may be due to an inadequacy of the grid used.

\section{Conclusion}
We have shown how a simple 1D Markov Process, the Zero Range Process, can be used to obtain rigorous results on the problem of 
parametrization of the passive scalar transport problem, relevant to many geophysical applications including temperature distribution in climate modeling. Using this model, we have derived rigorous results on the link between a principle of maximum entropy production and the principle of maximum Kolmogorov- Sinai entropy using a Markov model of the passive scalar diffusion called the Zero Range Process. 
 The Kolmogorov-Sinai entropy seen as function of  the convective velocity  admit a unique maximum. We show analytically that  both have the same Taylor expansion at the first order in the deviation from equilibrium. The behavior of these two maxima is explored as a function of the resolution $N$ (equivalent to the number of boxes, in the box approximation).  We found that  for a fixed difference of potential between the reservoirs , the maximal convective velocity predicted by the 
maximum entropy production principle tends towards a non-zero value, while the maximum predicted using Kolmogorov-Sinai entropy  tends to $0$ when $N$ goes to infinity. For values of $N$ typical of those adopted by  climatologists ($N \approx 10 \sim 100 $), we show that the two maxima nevertheless coincide even far from equilibrium. Finally, we  show that there is an optimal resolution $N_*$ such that the two maxima coincide to second order in $\bar{s}$,  a parameter proportional to the non-equilibrium fluxes imposed to the boundaries.
The fact that the optimal resolution depends on the intensity of the convective phenomena to be represented, points to new interesting research avenues, e.g. the introduction of convective models with adaptive grids optimized with maximum entropy principles on the basis of the convective phenomena to be represented.\\
On another hand, the application of this principle to passive scalar transport parametrization is therefore expected to provide both the value of the optimal  flux, and of the optimal number of degrees of freedom (resolution) to describe the system. It would be interesting to apply it to more realistic passive scalar transport problem, to see if it yield to model that can be numerically handled (i.e. corresponding to a number of bow that is small enough to be handled by present computers).
In view of applications to atmospheric convection, it would be interesting to apply this procedure to the case of an active scalar, coupled with a Navier-Stokes equation for the velocity. In such a case, the role of $f$ will be played by the turbulent subgrid Reynolds stresses. The heat fluxes and $N^*$ will be fixed by the coarse-graining length, and the optimization procedure will in principle provide the optimum subgrid Reynolds stresses at a given resolution $N$. Moreover, by imposing coincidence of MKS and  MEP,  one could get both the Reynolds stresses, heat fluxes and the optimum resolution.
Moreover, on a theoretical side, it will be interesting to study whether for general dynamical systems, there exists a smart way to coarse grain the  Kolmogorov- Sinai entropy  such that its properties coincide with the thermodynamic entropy production. This will eventually justify the use of the MEP principle and explain the deviations as well as  the different representations of it due to the dependence of the dynamic (Kolmogorov Smirnov, Tsallis, Jaynes)  entropies on the kind of partition adopted.

%\nocite{*}
 \bibliographystyle{Copernicus}
 \bibliography{biblio} 

\section{Appendix: Computation of the K-S entropy}
In this appendix, we compute the Kolmogorov-Sinai entropy  for the Zero Range Process, starting from its definition Eq. (\ref{eq:9}).
In the frame of our Zero Range Process , we use Eqs. (\ref{eq:9}) and (\ref{eq:proba2}) to write it as:

\begin{multline}
h_{KS}=-\sum_{i} \mu_{i_{stat}} \sum_{j} p_{ij}\log(p_{ij}) 
= -\sum_{m_1=0}^{+\infty}...\sum_{m_N=0}^{+\infty}
 P(m_1,m_2,...,m_N) \sum_{j} p_{(m_1,...,m_N) \rightarrow j} \log(p_{(m_1,...,m_N) \rightarrow j})\\
=-\sum_{m_1=0}^{+\infty} P(m_1)...\sum_{m_N=0}^{+\infty} P(m_N)
  \sum_{j} p_{(m_1,...,m_N) \rightarrow j} \log(p_{(m_1,...,m_N) \rightarrow j})
\end{multline}

We thus have to calculate $ \sum_{j} p_{(m_1,...,m_N) \rightarrow j} \log(p_{(m_1,...,m_N) \rightarrow j})$ that we will refer to as $(*)$ . We will take $p+q= \alpha+\delta=\beta+\gamma=1$ and $dt=\frac{1}{N}$  in order to neglect the probabilities to stay in the same state compare to the probabilities of changing state. There are five different cases to consider:

\begin{enumerate}

\item if $\forall i$ $  m_i \geq 1$  so the possible transitions are:\\
 $(m_1,m_2,...,m_N) \rightarrow (m_1 \pm 1,m_2,...,m_N)$ with respective probabilities  $ \alpha$ and $ \delta$\\
 $(m_1,m_2,...,m_N) \rightarrow (m_1 ,m_2,...,m_N \pm 1)$  with respective probabilities $ \gamma$ and $\beta$\\
 and $(m_1,...,m_k,...,m_N) \rightarrow (m_1,...,m_k \pm 1,...,m_N)$ with respective probabilities $p$ and $q$\\
\\
Thus,

\begin{equation}
(*)= \alpha \log \alpha +\delta \log \delta  +\gamma \log \gamma +\beta \log \beta + (N-1)(p\log(p)+q\log(q))
\end{equation}

\item if $m_1 \geq 1$ and $ m_N \geq 1$ and let $i$ be the number of $m_i$ between $2$ and $N-1$ equal to $0$. With the same argument as previously we have:

\begin{equation}
(*)= \alpha \log \alpha +\delta \log \delta  +\gamma \log \gamma +\beta \log \beta + (N-1-i)(p\log(p)+q\log(q))
\end{equation}

\item if $m_1 =0$ and $ m_N \geq 1$ and let  $i$ the number of $m_i$ between $2$ and $N-1$ equal to $0$ we have:

\begin{equation}
(*)= \alpha \log \alpha +\delta \log \delta   +\beta \log \beta + (N-2-i)p\log(p)+(N-1-i)q\log(q)
\end{equation}

\item The same applies if $m_1 \geq 1$ and $ m_N =0$ and let $i$ the number of $m_i$ between $2$ and $N-1$ equal to $0$ we have:

\begin{equation}
(*)= \alpha \log \alpha +\delta \log \delta   +\gamma \log \gamma + (N-1-i)p\log(p)+(N-2-i)q\log(q)
\end{equation}

\item finally, if $m_1 =0$ and $ m_N =0$ and let $i$ the number of $m_i$ between $2$ and $N-1$ equal to $0$ we have:

\begin{equation}
(*)= \alpha \log \alpha +\delta \log \delta  + (N-2-i)(p\log(p)+q\log(q)
\end{equation}

\end{enumerate}

Using equation \ref{eq:proba0}  we find that
$  P(m_k=0) =1- z_k$ and $ \sum_{i=1}^{+\infty} P(m_k=i) = z_k$ 
\\

We thus obtain than $ h_{KS}$ writes:

\begin{multline}
 h_{KS}= -(\alpha \log \alpha +\delta \log \delta  +\gamma \log \gamma +\beta \log \beta + (N-1)(p\log(p)+q\log(q)) \\
+(p\log(p)+q\log(q)) (\sum_{r=0}^{N} r \sum_{{i_1...i_N}} \prod_{i=i_1,...i_r}(1-z_i) \prod_{i \neq i_1...i_r}z_i)\\
+(\gamma\log(\gamma)+p\log(p))   z_N(1-z_1)(\sum_{{i_2...i_{N-1}}} \prod_{i=i_2,...i_r}(1-z_i) \prod_{i \neq i_2...i_r}z_i)\\
+(\beta\log(\beta)+q\log q)z_1(1-z_N)(\sum_{{i_2...i_{N-1}}} \prod_{i=i_2,...i_r}(1-z_i) \prod_{i \neq i_2...i_r}z_i)\\
+(\beta\log(\beta)+\gamma\log\gamma+p\log p+q\log q)(\sum_{{i_2...i_{N-1}}} \prod_{i=i_2,...i_r}(1-z_i) \prod_{i \neq i_2...i_r}z_i)
\end{multline}

This expression, though complicated at first sight, can be simplified. Indeed interested in the function $F(a)=\prod_{1}^{N} (z_k+a(1-z_k))$ and by deriving subject to $a$ we show that:
\begin{equation}
 \sum_{r=0}^{N} r \sum_{{i_1...i_{N}}} \prod_{i=i_1,...i_r}(1-z_i) \prod_{i \neq i_1...i_r}z_i=\sum_{i=1}^{N} (1-z_i)
\end{equation}

Thus we can simplify the last equation and we obtain:

\begin{multline}
\label{eq:hks3}
h_{KS}= -(\alpha \log \alpha +\delta \log \delta  +\gamma \log \gamma +\beta \log \beta + (N-1)(p\log(p)+q\log(q)))
+(p\log(p)+q\log(q)) \sum_{i=1}^{N} (1-z_i) \\
+(\gamma\log(\gamma)+p\log(p))(1-z_1) 
+(\beta\log(\beta)+q\log(q))(1-z_N) \\
\end{multline}

\end{document}